\documentstyle[12pt]{article}
\input epsf.tex
\begin{document}
\thispagestyle{empty}
 \noindent
\hspace*{115mm} hep-th/9607054 \\ 
\hspace*{115mm} TUW-96-11\\
\hspace*{115mm} July 1996\\
\begin{center} \vspace*{10mm} 
\Large \bf Comments on D-Brane and $SO(2N)$ Enhanced \\
\vspace*{5mm} Symmetry in Type II String\\ 
\vspace*{15mm} \large H. B. Gao$^\star$ \\
\normalsize{\it Institut f\"ur Theoretische Physik, TU Wien\\
Wiedner Hauptstr. 8-10, A-1040 Wien, Austria}\\
\vspace*{35mm} \large Abstract \\ \end{center}
We propose a configuration of D-branes welded by analogous
orbifold operation to be responsible for the enhancement
of $SO(2N)$ gauge symmetry in type II string compactified on the 
$D_n$-type singular limit of K3. Evidences are discussed from
the $D_n$-type ALE and D-manifold point of view. A subtlety
regarding the ability of seeing the enhanced  $SO(2N)$ gauge symmetry 
perturbatively is briefly addressed.  

\vspace*{30mm} 
\noindent\underline{~~~~~~~~~~~~~~~~}

\noindent $\star$ {\small On leave from Zhejiang University, China}
\newcommand{\be}{\begin{equation}}
\newcommand{\ee}{\end{equation}}
\newpage

It is well known that when N parallel D-branes coincide, the N copies
of the $U(1)$ gauge field get enhanced to the non-abelian $U(N)$ gauge
theory on the worldvolume\cite{wit1}. This has been employed in the context
of type II string compactified on the singular limit of K3 or C-Y to
explain enhancement of $A_n$ -type gauge symmetry when there are 2- or 
3-cyles on the manifold shrinking to zero or to a curve
\cite{dd}\cite{enh1}\cite{enh2}\cite{asp1}.
The work of \cite{dd} also interprets the appearance of the $U(N)\times U(M)$
gauge group as the result of two sets of D-branes intersecting on their
common transverse coordinates. As explained in \cite{asp2} and \cite{wit2},
enhanced gauge symmetries can appear only when the underlying conformal
field theories become singular. Indeed it was suggested in \cite{wit2} and
further demonstrated in \cite{ov} that at the moduli space where the
type IIA string compactified on the ALE manifold has $A_n$ type enhanced
gauge symmetry, the singular conformal field theory is such that it 
develops a semi-infinite tube similar to a symmetric fivebrane \cite{5bcft}
which admits exact (4,4) superconformal description. 

Little is still understood about the other examples of enhanced gauge
symmetry like $D_n$ and $E_{6,7,8}$ series which are believed to appear
when the (transverse) singularities are the corresponding type of ALE 
spaces. The work of \cite{ov} has suggested a general procedure to
study the ADE type singularity in the local models of K3 
compactification. It is essential that by performing discrete modding
of the local WZW coset at specific levels one obtains, through analog
of the stringy cosmic string construction \cite{cosm}, exact CFT of the
symmetric fivebrane with $H$-charge. Although \cite{ov} only worked
out details for the $A_n$ series, generalizations to other types of
ALE singular space are possible. We will deal with some delicacies
for the $D_n$ series in the following. Similar arguments involving
$T$- and $S$-duality \cite{dd} lead us to the D-string on D-manifold
(a manifold with a set of D-branes on it)
in the type IIB setting, where the enhanced gauge symmetry gets
explained in terms of the Chan-Paton factors of the open strings
stretched between the D-branes. In this letter we will propose a
non-compact D-manifold which consists of D-branes welded by an
analog of $Z_2$ orbifold operation dictated specifically by the
$D_n$ -type ALE space. Consideration of the compact D-manifold
has led to Vafa's F-theory \cite{vaf}. Our final picture seems to
be related to the dynamical D-branes in the type I string compactified
on the $Z_2$ orbifold limit of K3 \cite{gimpol}\cite{orient},
 though the differences are apparent, because 
we don't do orientifolding which would bring type IIB to type I,
and the ALE spaces are non-compact thus no tadpole concellation and
no 9-branes.  

That the D-manifold with two sets of $n$ D5-branes related by a $Z_2$
leads to $SO(2n)$ enhanced gauge group through Chan-Paton factors can
be roughly seen as follows. Suppose placing $n$ parallel D5-branes
at the point $y^i=a^i, i=1,..., n$ in the tranverse directions, 
and another
set of $n$ parallel D5-branes at the point $y^i=b^i, i=1,...,n$.
When these D5-branes coincide $a^i=a, b^i=b$, one gets for each sets
separately an enhanced group $U(n)$. Now if the two sets of branes are
such that their transverse coordinates are related by a $Z_2$ 
transformation $a^i=-b^i$, and the $U(1)$ gauge fields associated with 
individual D5-branes in different set are also identified by the induced 
$Z_2$ action, then, 
one gets the enhanced $SO(2n)$ gauge symmetry with massless adjoint
vector multiplets coming from open string states that would have given 
rise to $U(n)\times U(n)$ adjoint had the two sets of branes been 
arranged to intersect. Note that to obtain the above spectrum, it
is crucial that the open string states stretching from D5-branes in
one set to those in another won't be counted (this is the major
distinction from the type I orientifold), as their masses are
proportional to $T\vert a^i-b^j \vert \sim 2T\vert a \vert $ which is
not vanishing unless $a,b=0$. In the later case the configuration is
actually equivalent to $2n$ parallel D5-branes giving rise to $U(2n)$.
It is a degeneration of the $SO(2n)$ configuration and it is believed
that there are obstructions for the configuration to 
decay into $a,b=0$. So we will treat the generic case that $a,b\neq 0$
here. It should be mentioned that D-branes on the ALE space of type
$D_n$ must be more severely constrained than they are in the 
orbifold limits of K3, since the $D_n$ -type ALE space is essentially
a non-abelian orbifold. The relation $a^i=-b^i$ above should not
be taken as the $Z_2$ identification of the underlying orbifold,
rather it is a result of nontrivial dynamics of the D-branes moving
on the ALE space. The two sets of D-branes are being brought to the  
positions $a^i=-b^i$, but there are no $Z_2$ identifications of the
coordinates of the ALE space.

To give evidence supporting the above picture, we will study the
singular CFT of the $D_n$ -type ALE space and see if $D_n$ -type
D-manifold arise naturally. It is worthwhile repeating some logic 
steps outlined in \cite{dd}\cite{ov}.

Guided by the conifold in Calabi-Yau \cite{strom}, the work of
\cite{ov} advocates the role of 2d massless black holes in the
description of K3 and Calabi-Yau singularities, generalizing the
previous partial results in \cite{wit2}.
Starting from type IIA string compactified on singular limit of
K3, one models the local geometry of the singular K3 by the kind
of Landau-Ginzburg (orbifold) models which, by analogous arguments
of the $c=1$ string at self-dual radius, are of the form of the
(Gepner) product of two coset models $SL(2)/U(1)\times SU(2)/U(1)$
with relevant affine levels dictated by the ADE types of the minimal
model modular invariants and the requirement that the complete
models have ($N=2$) central charge $\hat{c}=2$. The non-compact
Kazama-Suzuki model $SL(2)/U(1)$ at the critical point is inspired
directly from the 2d Euclidean black hole which admits a semi-infinte
cigar as its bosonic geometry with a linear dilaton in the direction
of the length, and in addition, a circular scalar with 
radius associated with the affine level. By performing 
discrete group modding of the L-G superpotentials one obtains 
for the case of $A_{n-1}$ and  $G=Z_n$, a system equivalent to the
capped version of the symmetric fivebranes with $H$-charge $n$.
Implicit in the construction is a $T$-duality which exchanges the
$A$- and $B$-model descriptions and consequently type IIA
on ALE becomes type IIB on the symmetric fivebranes. 
Now type IIB in ten dimensions
admits $SL(2,Z)$ self-duality and therefore applying the $S$-duality
to the type IIB string on (NS-NS) fivebranes leads us to the
realm of D-string (the $S$-dual of the type IIB string) on the
D-5-branes (the $S$-dual of the NS-NS fivebranes). The authors of
\cite{dd} went on further to construct several examples of D-manifold
which consists of D-brane skeletons with open strings connecting
them. When these D-branes become coinciding, the massless states
which are now interpreted as excitations of the D-string appear
and give rise to vector as well as hypermulitplets of the extended
gauge theory on the worldvolume of the D-5-branes.  

We are interested here in seeing similar things happening in the 
case of $D_n$ series. At first glance, it seems difficult to conceive 
the appearance of the fivebrane CFT by applying directly
the discrete group modding to the L-G model of the $D_n$ type.
Actually the case of $D_n$ series corresponds to a non-Abelian orbifold
which seems less understood in the previous studies. At best we
could hope for an {\it orbifold} of the fivebrane CFT. This is
not really bad since we don't expect parallelly aligned {\it bona fide}
fivebranes to give rise to the $D_n$ type enhanced symmetry.
Keeping in mind the anticipation of the D-brane welding stated 
a while before, we are going to extrapolate a $Z_2$ action from the 
non-Abelian orbifold which should act on the fivebranes (and
subsequently on D-5-branes) as anticipated. 

For the type IIA string on $K3\times {\bf R}^6$ with the $D_n$ type 
singularity of K3
\be
D_n :~~ x^{n-1}+xy^2+z^2=0, ~~n\geq 4, \label{dn}
\ee
locally the Calabi-Yau geometry can be modelled by the Landau-Ginzburg
superpotential
\be 
W_{D_n}=-\mu w^{-2(n-1)}+x^{n-1}+xy^2+z^2. \label{dnw}
\ee
It can be verified that $W_{D_n}=0$ satifies the Calabi-Yau condition with
$\hat{c}=2$. The different parts of the terms in (\ref{dnw})
$(x,y,z)$ and $(w)$ are associated with the $N=2$ superconformal 
minimal model coset $SU(2)_{2(n-2)}/U(1)$ of type $D_n$ modular invariants
and the non-compact Kazama-Suzuki coset $SL(2,R)_{2n}/U(1)$ at level
$2n$, respectively. As in the $A_{n-1}$ case, one is obliged to perform
modding by a discrete group $G$ of the product of cosets.
Unlike the case of  $A_{n-1}$, here the discrete group $G$ happens to
be the rank $(n-2)$ dihedral group, ${\cal D}_{n-2}$ as it is the
symmetry of polynomial $W_{D_n}$ in eq(\ref{dnw}). 
Modding out the CFT by general (solvable)
discrete groups has been studied in \cite{dvvv},
and we will use some of their results in the following.\footnote{
Near the completion of this work, we became aware of the references
\cite{zas1, zas2} where non-Abelian orbifolds of $CP^1$ and  $CP^2$
by dihedral groups have been studied in connection with cohomology
rings of the $\sigma$-model target spaces.}

The group ${\cal D}_{n-2}$ can be taken as semi-direct product of $Z_2$
and $Z_{n-2}$ and is defined 
by two generators $\theta, \tau$ satisfying relations 
\be
\tau^2=\theta^{n-2}=1; ~~\tau^{-1}\theta \tau=\theta^{-1}.
\ee
Generator $\theta$ corresponds to rotation of the $(n-2)$-sided polygon
through an angle $2\pi/(n-2)$, while generator $\tau$ a rotation of
$\pi$ about an axis of symmetry. The stabilizer subgroups are 
$N_1={\cal D}_{n-2}, ~N_{\tau}=Z_2, ~N_{\theta}=Z_{n-2}$. There are
two or four 1-dimensional irreducible representations and $\frac{1}{2}
(n-3)$ or $\frac{1}{2}(n-4)$ 2-dimensional irreducible representations
according to whether $n$ is odd or even.
The 1-dimensional ones act simply as sign-changes while the 2-dimensional
ones are generically of the form
\be
 \sigma_i(\tau)=\pmatrix{ 0 & 1 \cr 1 & 0 \cr}, ~~~
\sigma_i(\theta) =\pmatrix{\alpha^i & o \cr 0 & \alpha^{-i} \cr},~~~
i=1,..., l, \label{2drep}
\ee
where $\alpha$ is the $(n-2)$-th root of unity, and
$l$ is the number of the 2-dimensional representations in either
cases. The two dimensional representation (\ref{2drep}) is helpful
when we notice that the WZW models of $SL(2,R)$ and $SU(2)$ both
admit natural actions of ${\cal D}_{n-2}\subset SU(2)$.
The detailed spectrum and operator content of this
(non-abelian) orbifold conformal field theory can be worked out,
according to  \cite{dvvv}, at least for the cases of solvable
groups. We are not going to bring the burden of a complete
analysis of various sectors of chiral rings into the present work, 
but only note that in the fixed-point-free\footnote{Fixed-point
in the sense of fusion algebra, not to be confused with the 
geometric sense of ALE.} case, the linear  (i.e.,
not projective) representations in the untwisted sector give rise 
to operators which are in 1-1 correspondence with the nodes of
the extended Dynkin diagram. We ignore the twisted sectors and
the related fusion rules since we are not concerned with 
solving the orbifold CFT completely.

An essential point of the non-Abelian orbifolds studied in 
\cite{dvvv} is that since
the group ${\cal D}_{n-2}$ is (super)solvable, which means that
there is an exact sequence
\be 0 \rightarrow Z_2 \rightarrow {\cal D}_{n-2} \rightarrow
Z_{n-2} \rightarrow 0 \ee
such that ${\cal D}_{n-2}$ is the extension of $Z_{n-2}$
by $Z_2$, one can perform the orbifolding by ${\cal D}_{n-2}$ through
a sequence of Abelian orbifolds by $Z_{n-2}$ and $Z_2$.
If one performs first modding out by $Z_{n-2}$ and subsequently
performs modding by $Z_2$, the action of the $Z_2$ is to 
identify the operators obtained by conjugating the representations
in each sector of the twisted Hilbert space. However, we note that
by modding out $Z_{n-2}$ we always get trivial bosonic part of
the WZW $SU(2)$ model with its free scalar field at radius $1$. 
Thus it seems impossible to see the fivebrane CFT that is a
WZW $SU(2)$ model at nontrivial level and a Feigin-Fuchs field.
Examining the affine levels of both factors of the product coset,
one sees that if one scales the momentum lattice of the free bosons by a
factor of $\frac{1}{\sqrt{2}}$, then modding out by a cyclic group 
$Z_{n-1}$ would result in a WZW model $SU(2)$ at level $n-2$
\footnote{The effect of modding out $Z_{n-1}$ is to recombine
two bosons from the product WZW cosets into 
$$ \tilde{X}^1=p_1 X^1 -p_2 X^2 $$ $$  \tilde{X}^2=
\sqrt{(n-2)}p_1 X^1 + X^2$$
with radii $ \tilde{R}_1 =1, \tilde{R}_2=\sqrt{2(n-2)}$.} together
with a Feigin-Fuchs field which should correspond to a version of
symmetric fivebrane with the $H$-charge $n$. 

This seems to be what we want. But how could we modd by  $Z_{n-1}$ 
for a ${\cal D}_{n-2}$ system?  $Z_{n-1}$ is the group of 
monodromies of the $D_n$ singularity (see for example \cite{briesk}),
modulo the $Z_2$ center (i.e. the full-fledged monodromy group is 
$Z_{2(n-1)}$ instead of $Z_{n-1}$). By modding out by $Z_{n-1}$
we did nothing wrong except trivialized the monodromy around the $y$
coordinate in eq.(1). Incidently, letting $y=0$, one recovers, up to
quadratic perturbation, an $A_{n-2}$-type singularity, hence 
orbifolding by $Z_{n-1}$ (and letting $w'=w^2$ in eq.(2))gives CFT of 
fivebranes of $H$-charge $n-1$,
as in \cite{ov}. Compare to the result in the last paragraph, we
see an extra fivebrane has been trapped in the trivialization of the
$y$-monodromy.

Another point in the above argument is the effect of rescaling the 
momentum of the free boson by a factor of $\frac{1}{\sqrt{2}}$.
What does it do to the fivebrane picture? Note that this has 
effectively changed the K\"ahler moduli parameter. If we identify
the circle as a homology circle in the torus fiber of the ALE
space (after modelling it by analogous stringy cosmic string
construction), this rescaling  corresponds to modifying the
$B$ field period by one half, $B\rightarrow B + \pi$. The $H$-charge
$n$ which we got before actually consists of contributions of $2n$
symmetric fivebranes each carrying $H$-charge $\frac{1}{2}$.

Now it is clear how the remaining orbifold by $Z_2$ act on the 
$2n$ fivebranes. There are $(n-1)$ pairs of them identified by the
$Z_2$ as inner automorphism of the $A_{n-2}$ extended Dynkin diagram;
the other two are exchanged via the outer automorphism $Z_2$ which
effectively splits off the fivebrane trapped when trivializing the
$y$-monodromy. It would be a challenge to verify this explicitly by 
studying the inverse question of orbifolding the CFT of the $2n$
symmetric fivebranes
and deriving the $D_n$-type ALE geometry
as an object whose affine coordinate ring coincides with the
truncated chiral ring of the orbifold CFT.

Having exploited the relevant features of the (non-abelian) orbifold,
 we would like to further pursue the idea that
D-string on D-manifold of the $D_n$-type indeed gives rise to 
enhanced gauge symmetry $SO(2n)$. 
Now we have a picture of type IIB on the orbifold of the fivebranes
with $H$-charge $n$, according to \cite{dd}, the $S$-duality of the
10 dimensional type IIB string theory leads to the dual picture of
D-string on a manifold with D-branes. There is a subtlety though,
that is when we transform the $B$ field by the $SL(2,Z)$, we would
still get $n$ D-5-branes. The point is that the group
 $SL(2,Z)$ has a mod 2 congruence subgroup
$\Gamma_0(2) \subset SL(2,Z)$, which upon conjugation by the element
$\delta=\pmatrix{2 & 0 \cr 0 & 1 \cr} \in SL(2,Z)$ dictates the
kind of transformation of the $B$ field we found earlier. Thus if
we use $SL(2,Z)/ \delta \circ \Gamma_0(2)$ as the $Z_2$ part of the
$S$-duality group of the type IIB string theory, then obviously
the NS-NS fivebranes of charge $n$ get transformed into $2n$ D-5-branes. 

To see the enhanced gauge symmetry of type $SO(2n)$ in type IIA
string compactified on $D_n$ -type sigular limit of K3, we are
looking for the relevant D-manifold on which the dual of the 
type IIB string get excitations of the form of the
corresponding gauge multiplets.
It is straightforward to generalize the arguments of \cite{dd}
to the $D_n$ -type D-manifold, and find out the massless 
spectrum of the $N=2$ $SO(2n)$ gauge theory.
 
Let us see how the pure gauge content arise simply by
looking at what the D-manifold looks like.
For the case of $A_{n-1}$ series, the D-manifold coincides with
the dual of the corresponding extended Dynkin diagram
\footnote{The extended Dynkin diagrams of the simply laced
Lie algebras with specific categories are named quivers\cite{
nak}. These quiver diagrams contain more information than
the extended Dynkin diagrams and are quite useful in studying
ALE instantons via ADHM construction. See \cite{kro}\cite{more}
for the discussion of this from mathematical and physical 
points of view.}  (see fig. 1
of the paper \cite{dd}), where an extra vertex is added which
is expressable in terms of the $(n-1)$ simple roots of $SU(n)$
$$ r_0=-\sum^{n-1}_{i=1} r_i $$
where $r_i=\mu_i-\mu_{i+1}, ~i=1,... n-1$ are the simple roots.
Note that when $\mu_i=\mu_{i+1}$, signalling vanishing of the
the $(n-1)$ 2-cylces, the extra cycle vanishes trivially (as 
dictated by homology relation), though it is interpreted as
appearance of massless hypermultiplets in the adjoint.
The D-manifold dual to the extended Dynkin diagram can be viewed
as representing $n$ D-branes located at different positions
with links connecting them being identified as masses of the
stretched string states, which go to zero when the D-branes 
coincide. The extended Dynkin diagram of the $D_n$ type consists of
$n+1$ vertices, the $(n+1)^{th}$ vertex is defined as 
\be r_0=-\sum^n_{i=1} a_i r_i\ee
where $r_i$ denote 
%$a_{n-2}=2, ~a_i=1, ~i\neq n-2$, and 
the simple roots of $SO(2n)$,
$\mu_1-\mu_2, ..., \mu_{n-1}-\mu_n, \mu_{n-1}+\mu_n$, and
$a_i$ are the dimensions of the irreducible representations of
the groups ${\cal D}_{n-2}$. We recall that the number of
two dimensional characters of ${\cal D}_{n-2}$ is $\frac{n-4}{2}$
or $\frac{n-3}{2}$ depending on whether $n$ is even or odd.
we note that vanishing of the $n$ 2-cycles must be accompanied with
a $Z_2$ symmetry of reflection on the $\mu_n$ plane. This 
is part of the $Z_2$ symmetry belonging to 
the Weyl group of $SO(2n)$ 
root system. Now it is clear where to put our $2n$ D-branes
on the (resolved) ALE space $X_{ALE}=
{\bf C}^2/{\cal D}_{n-2}$. 
We identify the root lattice of the simply-laced group
with the second homology group $H_2(X_{ALE},Z)$ of the
(resolved) ALE space $X_{ALE}$, the vertices of the Dynkin
diagram can be taken as the vanishing 2-cycles whose areas
are proportional to the distance of the nearby D-branes
$\mu_i -\mu_{i+1}$. We imbed the root lattice of the
two copies of the $A_{n-2}$ system into that of the $D_n$
system. According to the previous discussions,
the $2(n-1)$ D-branes are identified pairwise by the
inner automorphism $Z_2$ of the $A_{n-2}$ system as
(pairwise) permutations of the vectors $\mu_i, i=1,..., n-1$.
There are also two D-branes which are exchanged by the
outer automorphism (of the $A_{n-1}$ system), here represented 
by the Weyl reflection on the vector $\mu_n$. The
dual picture of this $D_n$ -type D-skeletons is depicted as
the extended Dynkin diagram in Fig. 1 below (with $n=6$)
\vskip 1cm
\hskip1.5cm\epsfxsize=4.0in\epsfbox{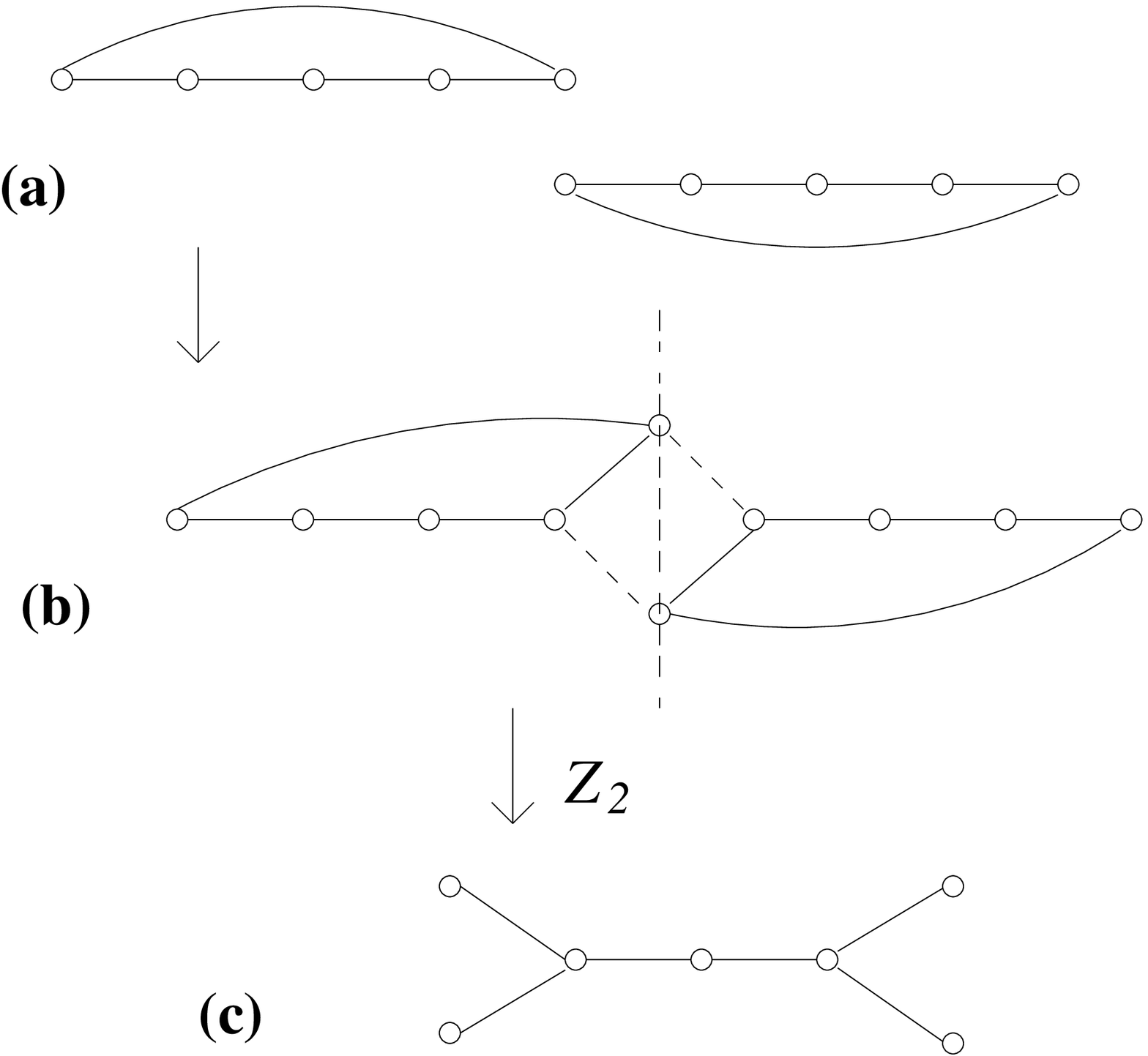}
\vskip0.4cm
\centerline{{\bf fig 1.} \small A sequence of steps for obtaining
$D_n$ extended Dynkin diagram from two $A_{n-2}$'s.}
\vskip 0.4cm
 In Fig.1b, one sees the hidden D-brane between the
$(n-2)^{th}$ vertex and another vertex which is the image of the
$(n-1)^{th}$ vertex under outer automorphism. 

Before closing, we mention that there is a subtlety in the
enhanced gaueg symmetry of the $D_n$ series. Namely, one might ask
whether this kind of enhanced gauge symmetry can be seen 
perturbatively in the heterotic duals
or is it intrinsically related to some non-perturbative
effects like small instantons in the heterotic $SO(32)$ theory
\cite{wit3}. There is indication that the vertex with three links
ending on it should correspond to the 2-cycle whose vanishing can
not be understood perturbatively \cite{asp1}. In our case an
outer automorphism acts nontrivially on this vertex, it must
be checked whether this action is compatible with (possible)
monodromy actions when fibering the $D_n$ ALE space over a base.
This and other questions related to string-string duality
are certainly worth pursuing further. 

\noindent {\large \bf Acknowledgements}

We are grateful to A. Brandhuber and M. Kreuzer for discussions.
This work is supported by the FWF under project no. P10641-PHY.

\eject

\end{document}